\documentclass[a4paper,conference]{IEEEtran}

\usepackage{graphicx}
\graphicspath{ {./images/} }

\usepackage{amsmath}
\usepackage{xcolor}

\begin{document}

\title{Dynamic Background Subtraction\\ by Generative Neural Networks}

\author{\IEEEauthorblockN{Fateme Bahri}
\IEEEauthorblockA{Department of Computing Science\\
University of Alberta, Canada\\
Email: fbahri@ualberta.ca}
\and
\IEEEauthorblockN{Nilanjan Ray}
\IEEEauthorblockA{Department of Computing Science\\
University of Alberta, Canada\\
Email: nray1@ualberta.ca}}

\maketitle

\begin{abstract}
Background subtraction is a significant task in computer vision and an essential step for many real world applications. One of the challenges for background subtraction methods is dynamic background, which constitute stochastic movements in some parts of the background. In this paper, we have proposed a new background subtraction method, called DBSGen, which uses two generative neural networks, one for dynamic motion removal and another for background generation. At the end, the foreground moving objects are obtained by a pixel-wise distance threshold based on a dynamic entropy map. The proposed method has a unified framework that can be optimized in an end-to-end and unsupervised fashion. The performance of the method is evaluated over dynamic background sequences and it outperforms most of state-of-the-art methods. Our code is publicly available at https://github.com/FatemeBahri/DBSGen.
\end{abstract}

\IEEEpeerreviewmaketitle

\section{Introduction} \label{sec:Introduction}
Background subtraction is an effective approach for change detection problem that is a fundamental task in computer vision applications, such as video surveillance, autonomous navigation, traffic monitoring and Human computer interaction \cite{bouwmans2019deep,garcia2020background}. Different methods have been proposed for background subtraction, however many of these methods are vulnerable to image sequences with dynamic background. In a scene with dynamic background, a background pixel can have different values due to periodical or irregular movements \cite{xu2016background}. For example, surging of water, water fountains and waving trees cause dynamic variations in the background. Segmenting such dynamic background variations from foreground is a challenging task and negatively affects the methods' performance.

In background subtraction, methods aim to acquire a background model (BM) in order to segment moving objects and foreground from an input image. One traditional group of methods is based on probability density estimation by observing pixel value statistics. In statistical parametric methods, usually Gaussian functions are used to model the pixel values in a sequence. A single Gaussian model was proposed in \cite{wren1997pfinder}, however, a single function was not enough to model variations in pixels' history. Later, Gaussian mixture model (GMM) \cite{stauffer1999adaptive}, a traditional and still a popular method was proposed that models pixels by a number of Gaussians. Afterwards, improved variations of GMM were introduced in \cite{zivkovic2004improved, zivkovic2006efficient, lee2005effective, kaewtrakulpong2002improved} to enhance the method. The parametric methods may be vulnerable to sudden changes in the scene. To address this issue, a statistical non-parametric algorithm, KDE \cite{elgammal2000non}, was presented that estimates probability of pixel values using kernel density estimation.

A major category of methods utilize controller parameters to update BM based on dynamic feedback mechanisms. SuBSENSE method \cite{st2014subsense} makes use of color channels intensity and spatio-temporal binary features. In addition, it dynamically tunes its parameters by pixel-wise feedback loops based on segmentation noise. PAWCS \cite{st2015self}, one of the state-of-the-art methods, is an extended version of SuBSENSE that generates a persistent and robust dictionary model based on spatio-temporal features and color. Like SuBSENSE, PAWCS automatically adjust itself using feedback mechanisms. SWCD method \cite{isik2018swcd} utilizes dynamic controllers of SuBSENSE in addition to a sliding window approach for updating background frames. CVABS \cite{icsik2019cvabs}, is a recent subspace-based method which employs dynamic self-adjustment mechanisms like SuBSENSE and PAWCS.

A new category of algorithms for change detection are ensemble methods. Recently, In \cite{bianco2017combination,bianco2017far}, authors proposed a few versions of a method called IUTIS (In Unity There Is Strength) that exploits genetic programming (GP) to combine various algorithms to leverage strength of each. GP selects the best methods, combine them in different ways and applies right post-processing techniques. IUTIS combines several top-ranked methods evaluated on CDnet 2014 (\cite{wang2014cdnet}) and it achieves a good performance.

In recent years, numerous methods based on deep neural networks (NN) have been proposed due to success of deep learning in computer vision applications. Foreground Segmentation Network (FgSegNet) and its following variations \cite{gao2021extracting,lim2018foreground,lim2020learning} are currently the state-of-the-art based on their performance on CDnet 2014. Motion U-Net \cite{rahmon2021motion} is another deep NN method and needs less parameters than FgSegNet. BSPVGAN \cite{zheng2020novel} uses Bayesian Generative Adversarial Networks (GANs) to build the background subtraction model. Cascade CNN \cite{wang2017interactive} is another method that employs a multi-resolution convolutional neural network (CNN) for segmenting moving objects. DeepBS \cite{babaee2018deep} trains a CNN with patches of input images and then merge them to rebuild the frame; It utilizes temporal and spatial median filtering to improve the segmentation results. Another supervised method called BSUV-Net \cite{tezcan2020bsuv,tezcan2021bsuv} trains on some videos and their spatio-temporal data augmentations. After training, BSUV-Net can perform well on unseen videos. The mentioned NN methods are top-ranked among the evaluated methods on CDnet 2014. However, they need supervised training, meaning they require pixel-wise annotated ground-truth that is an expensive manual task and not practical in every situation.

Some of the recent proposed methods, SemanticBGS \cite{braham2017semantic} and its variations RT-SBS-v1 and RT-SBS-v2  \cite{cioppa2020real}, combine semantic segmentation with background subtraction algorithms. They leverage the information of a semantic segmentation algorithm to obtain a pixel-wise probability to improve the output result of any background subtraction algorithms. We do not consider them for comparison because they are using a pixel-wise information as input even though they do not get trained by ground-truth labels.

The top-ranked reported methods on CDnet website that do not have supervised learning or using any other pixel-wise input information are PAWCS \cite{st2015self}, FTSG \cite{wang2014static}, SWCD \cite{isik2018swcd} and CVABS \cite{icsik2019cvabs} methods. FTSG (Flux Tensor with Split Gaussian models) runs flux tensor-based motion segmentation and a GMM-based background modeling separately, then fuses the results. At the end, it enhances the results by a multi-cue appearance comparison.

\begin{figure*}[ht!]
\centering
\includegraphics[width=6in]{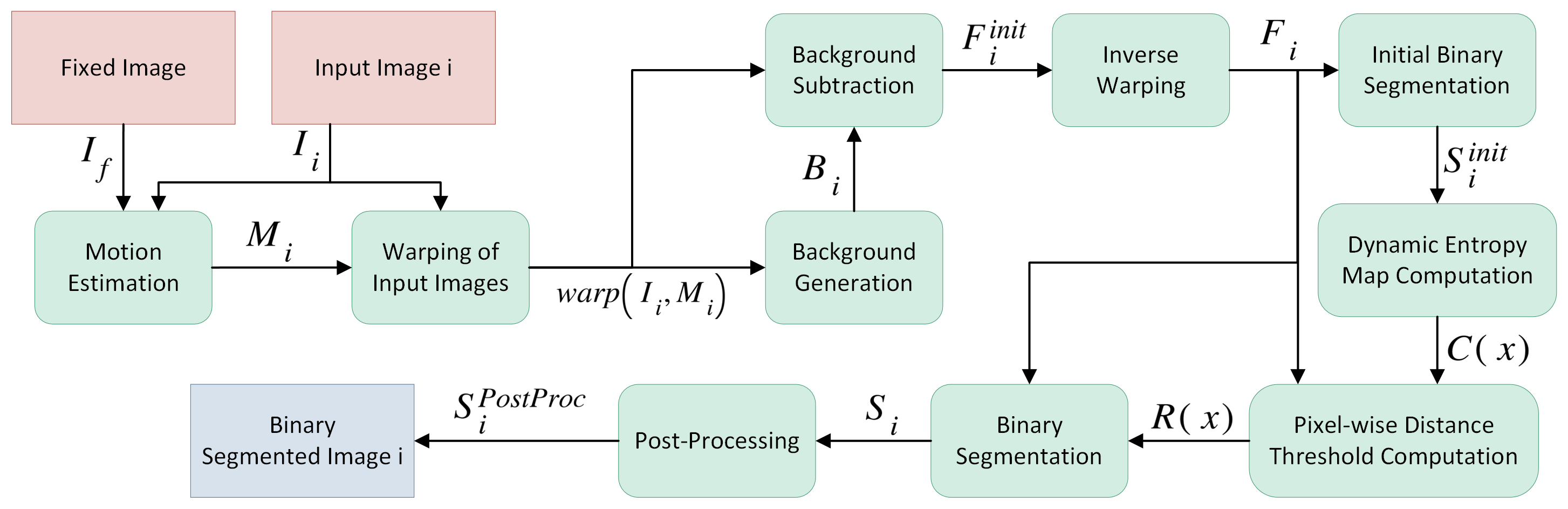}
\caption{Block diagram of DBSGen}
\label{fig:Block_diagram}
\end{figure*}

In this paper, we have proposed a Dynamic Background Subtraction by Generative neural networks (DBSGen). DBSGen exploits a generative multi-resolution convolutional network to estimate a dense motion map that minimizes the difference between each input image and a fixed image. The fixed image is chosen from the video as an initial background model. Next, our method warps each input image using its pixel-wise motion map. In the warped images, most of pixels due to the dynamic motions are mapped to pixels of the fixed image. However, some moving objects are also warped in the process. Subsequently, DBSGen leverages a generative fully connected network \cite{bahri2018online} to generate background images for the warped input images. Then, foreground images are obtained by subtracting background images from warped images. Afterwards, an inverse warping of the motion map is applied on the foreground images to warp back the moving objects, otherwise, results would contain deformed objects. Then, inspired by SuBSENSE method \cite{st2014subsense}, DBSGen computes a pixel-wise dynamic entropy map that is an indicator of dynamic background spots. By utilizing this map, a pixel-wise distance threshold is achieved. Next, DBSGen obtains binary segmented images using the distance threshold. Finally, some basic post-processing operations enhance the results. A block diagram of DBSGen is presented in Fig. \ref{fig:Block_diagram}.

DBSGen's contributions can be summarized as follows. First, it estimates a pixel-wise motion map by a generative network and exploits it for dynamic background subtraction problem. Second, unlike many other neural network based methods, it is optimized in an unsupervised way, without requiring expensive pixel-wise ground-truth masks. Third, it is an end-to-end neural network framework, which is optimized in one stage.  

The rest of the paper is organized as follows. Section \ref{sec:Method} explains details of DBSGen framework and how it performs dynamic background subtraction. In section \ref{sec:Results}, we report our implementation details, experimental results and comparison with state-of-the-art methods. Finally, Section \ref{sec:Conclusion} provides conclusions and an outline of the future work.

\section{Proposed Method} \label{sec:Method}
DBSGen is based on dynamic motion removal, background generation and pixel-wise thresholding. Optimizations of the networks are performed in an end-to-end manner. In the following subsections, the description of each of these steps is given.

\subsection{Motion Estimation}
\begin{figure*}[!ht]
\centering
\includegraphics[width=6in]{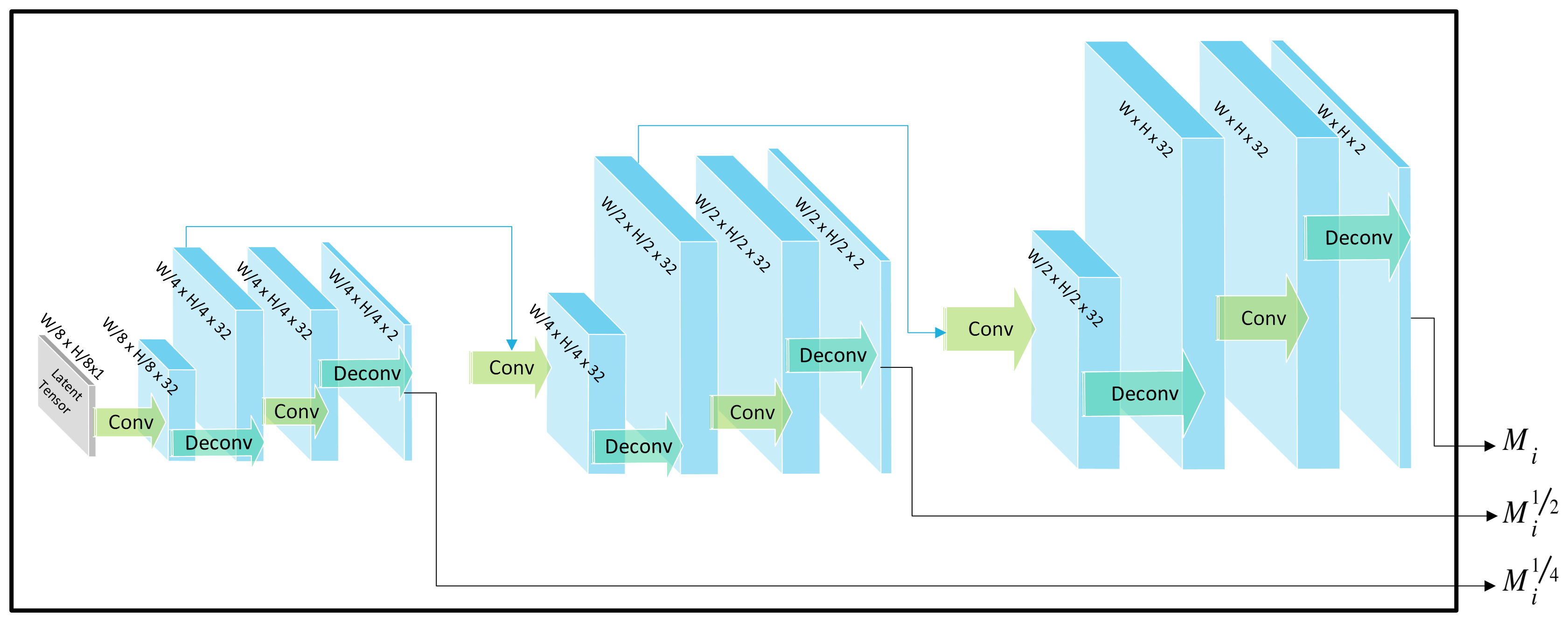}
\caption{Architecture of the Generative Multi-resolution Convolutional Network (GMCN). Input to the network is an optimizable latent tensor. Outputs are dense motion maps in three resolutions.}
\label{fig:GMCN}
\end{figure*}

By estimating a pixel-wise motion map, DBSGen aims to warp each input image such that it becomes similar to a fixed image. It helps to remove some of the dynamic background motions in the warped input images. For this purpose, we use a Generative Multi-resolution Convolutional Network (GMCN) that is inspired by \cite{sheikhjafari2018unsupervised}. It generates motion maps in three resolutions, from coarse to fine. We utilize it for estimating small motions including dynamic background motions in the input frames by applying a motion compensation loss. Fig. \ref{fig:GMCN} shows the GMCN's architecture.

The input to GMCN is an optimizable latent tensor with size $N \times H/8\times W/8$, where $N$ is the number of the frames in the sequence and $H$ and $W$ are the height and width of each image, respectively. GMCN computes 2D motion estimation maps in three resolutions called $M_i^{1/4}$, $M_i^{1/2}$ and $M_i$ that are used to warp the $i^{\text{th}}$ input frame of the sequence, $I_i$, and reduce dynamic background motions. The upsampled motion map of each resolution is added to the the higher resolution's motion map to refine it. In each resolution, a loss term is responsible for minimizing difference of the warped input frame and the fixed image. $L_{motion}$ loss term, optimizes parameters of the GMCN. 

\begin{equation} \label{loss_motion}
\begin{split}
&L^{res1/4} = \sum_{i=1}^N\Vert warp(I_i^{1/4}),M_i^{1/4}) - I_f^{1/4} \Vert_2,\\
&L^{res1/2} = \sum_{i=1}^N\Vert warp(I_i^{1/2},M_i^{1/2}) - I_f^{1/2} \Vert_2,\\
&L^{res1} = \sum_{i=1}^N\Vert warp(I_i,M_{i}) - I_{f} \Vert_2,\\
&L_{motion-reg} = \sum_{i=1}^N\Vert M_i^{1/4}\Vert_2 + \Vert M_i^{1/2}\Vert_2 + \Vert M_{i}\Vert_2 ,\\
&L_{motion} = L^{res1/4}+\lambda L^{res1/2}+\lambda^2L^{res1}+L_{motion-reg},
\end{split}
\end{equation}

where $\Vert.\Vert_2$ denotes the $L_2$-norm and $I_{f}$ represents a background image selected from one or an average of a few frames without a moving objects from the input sequence. Function $warp(I_i,M_i)$ warps the image $I_i$ with the pixel-wise motion map $M_i$. $\lambda$ is a hyper-parameter to control relative importance of the terms and its value is chosen by experiments. $L_{motion-reg}$ is a regularization term for motion maps that does not allow estimated motion values grow large. Although, we do it to avoid warping of moving objects still some motions of foreground moving objects are captured in the motion map and as a result, they get warped. That is why DBSGen applies an inverse warping, based on motion maps, on foreground images, in a later step.

\subsection{Background Generation}
Background is generated by a Generative Fully Connected Network (GFCN) that was proposed in our previous method called NUMOD \cite{bahri2018online}. GFCN has an optimizable low-dimensional latent vector as the input. The input layer is followed by three fully connected hidden layers each connected to a batch normalization layer. The activation functions of the first two layers are ELU \cite{clevert2015fast} and the last one is Sigmoid to limit output values between zero and one.

$L_{recons}$ loss term that is responsible for constructing background images is as follows:
\begin{equation} \label{loss_Reconst}
L_{recons} =\sum_{i=1}^N\Vert warp(I_i,M_i) - B_i \Vert_1 ,
\end{equation}
where $B_i$ is the $i^{\text{th}}$ output of GFCN and $M_i$ is the obtained motion map from GMCN. $\Vert.\Vert_1$ denotes the $L_1$-norm. We used $L_1$-norm instead of $L_2$-norm in $L_{recons}$ because it encourages sparsity \cite{candes2008enhancing}. 

\begin{figure*}[ht!]
\centering
\includegraphics[width=5.5in]{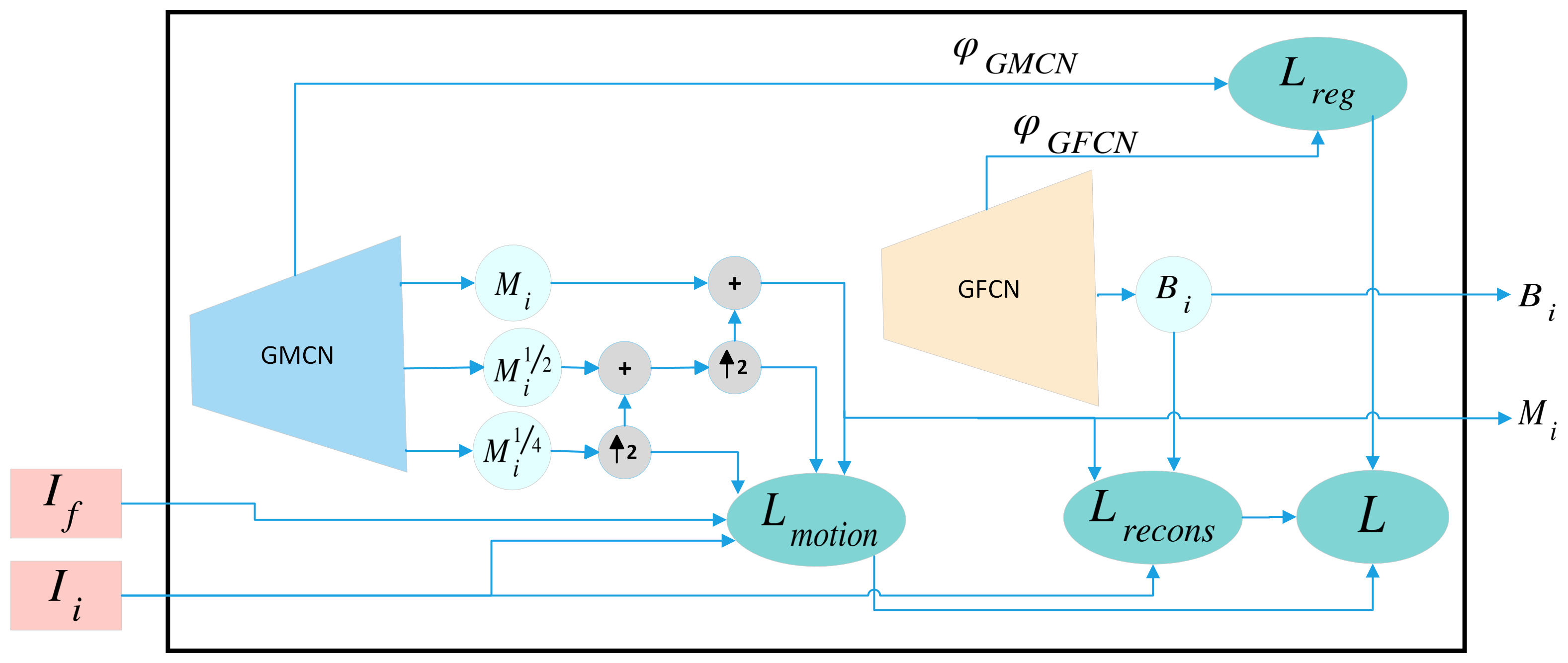}
\caption{The diagram shows the flow of the computations in our framework for the $i^{\text{th}}$ image frame. Input image $I_i$ and fixed image $I_f$ are shown in pink. GMCN, Generative Multi-resolution Convolutional Network, with optimizable parameters $\varphi_{GMCN}$, computes dynamic motion estimation maps. GFCN, Generative Fully Connected Network, with optimizable parameters $\varphi_{GFCN}$, generates the background image. Inputs to GMCN and GFCN are internal optimizable latent parameters. Outputs are the background image, $B_i$, and the dense motion map, $M_i$, for warping every pixel of $I_i$.}
\label{fig:overview}
\end{figure*}

GFCN behaves like a decoder in an autoencoder network with the difference that here, the input to to the decoder is an optimizable latent vector, which can learn a low-dimensional manifold of the data distribution by applying some constraints like limiting the capacity of the network and choosing a small input latent vector size ~\cite{goodfellow2016deep}. Since The network is able to extract the most salient features of the data and $L_{recons}$ loss term is imposing similarity of output and input frames, therefore, during optimization, GFCN learns a background model. This happens because the sequence of input images are temporally correlated to each other and the background part of images are common among them \cite{bahri2018online}.
The overall loss function of DBSGen is defined as:

\begin{equation} \label{loss}
L  = \alpha L_{recons} + L_{motion} + L_{reg},
\end{equation}
where $L_{reg}$ is the $L_2$ regularization that we apply on parameters of the networks to prevent overfitting to noise. $\alpha$ is a hyper-parameters to take into account relative importance of $L_{recons}$ term and is determined by conducting experiments.
The computation flow of DBSGen is shown in Fig. \ref{fig:overview} 

\subsection{Foreground Detection}
\begin{figure}[ht!]
\centering
\includegraphics[width=3in]{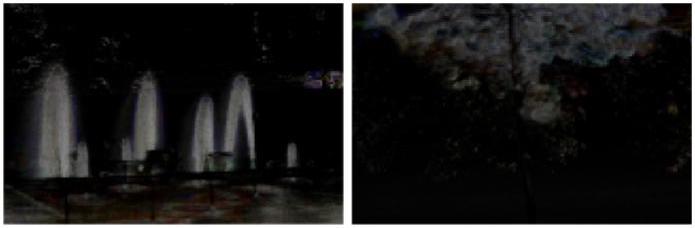}
\caption{Dynamic entropy map, $C(x)$, of ``fountain01'' and ``fall'' videos}
\label{fig:C_x}
\end{figure}

For obtaining foreground part of the images, $F_i^{init}$, our method subtracts the obtained background image from the warped input image. Then, it applies an inverse warping on the result to warp the moving objects back to their original shape and acquires foreground, $F_i$ as follows:
\begin{equation} \label{eq_F}
\begin{split}
&F_i^{init} = warp(I_i,M_i) - B_i, \\
&F_i = warp^{inverse}(F_i^{init},M_i).
\end{split}
\end{equation}
For obtaining the foreground mask, we use a pixel-wise thresholding method. This is adopted from SuBSENSE method \cite{st2014subsense} for detecting blinking pixels by measuring the dynamic entropy of each pixel. $C(x)$, dynamic entropy map, counts the number of times a pixel switches from being a foreground to a background or vice versa between consequent frames and is computed as follows: 
\begin{equation} \label{eq_counter}
C(x) =\frac{1}{N-1}\sum_{i=2}^N XOR (S_i^{init}(x),S_{i-1}^{init}(x)),
\end{equation}
where $x$ is a pixel and $S_i^{init}$ is the binary result of the $i^{\text{th}}$ frame in the sequence after an initial segmentation. This initial segmentation uses the standard deviation of all foreground frames, $F$, in each color channel as the distance threshold. Note that these three threshold values for RGB channels are same among all frames. Values of dynamic entropy map, $C$, are in the range $[0,1]$, where dynamic background regions would have greater values, while static background regions would have $C(x)\approx0$. Dynamic entropy map of ``fountain01'' and ``fall'' videos can be observed in Fig. \ref{fig:C_x}.

In the following step, we compute the pixel-wise distance thresholds:
\begin{equation} \label{eq_thresholds}
R(x) = \mu_{ch}+\beta_1\sigma_{ch}+\beta_2\sigma_{ch}C(x)+\beta_3\sigma^2_{C_{ch}}C(x),
\end{equation}

where $\mu_{ch}$ and $\sigma_{ch}$ are the mean and standard deviation of the foreground frames $F$ in each color channel, respectively, and $\sigma^2_{C_{ch}}C(x)$ is the variance of the counter $C$ in each color channel.
The binary segmented result, $S_i$, is obtained by applying $R(x)$ distance threshold on the foreground $F_i(x)$. 

Our post-processing step is minimal like other state-of-the-art methods \cite{st2014subsense,mandal2018candid}: we apply a median blur filter and binary morphological closing on $S_i$ to eliminate salt-and-pepper noise. The final binary segmented result is called $S_i^{PostProc}$.

\section{Experimental Results and Discussion}\label{sec:Results}

\subsection{implementation details}
DBSGen is implemented in TensorFlow platform. GFCN has an optimizable vector of size $3$ as its input and three fully connected layers of sizes $12$, $24$, and $43$, successively. Convolutional and deconvolutional layers in GMCN each have $32$ filters of size $7\times7$. Values of hyper-parameters $\lambda$ and $\alpha$ are set to $0.25$ and $0.1$, respectively, by conducting several trial and error experiments. Adam \cite{kingma2014adam} with learning rate of $0.006$ is used as the optimization algorithm. The whole framework is optimized in $50$ epochs in an end-to-end fashion. The average running time of DBSGen on Dynamic Background category of CDnet 2014 \cite{wang2014cdnet} is about 0.69 second per frame on a GeForce GTX 1080 Ti GPU.

\subsection{Dataset and Evaluation Metric}
We evaluate DBSGen on videos of Dynamic Background category of change detection (CDnet 2014) dataset \cite{wang2014cdnet} to validate its effectiveness in challenging dynamic background scenarios. It includes six videos; ``fountain01'' and ``fountain02'' contain dynamic water background, also, ``canoe'' and ``boats'' videos exhibit water surface motion, while ``overpass'' and ``fall'' videos have waving trees in their background. Due to the lack of space in Table \ref{table:motion_comparison}, we mention the videos with the following names: `fnt1'', `fnt2'', ``canoe'', ``boats'', ``over'' and ``fall''.

For evaluation, we use F-Measure (FM) metric that is used generally as an overall performance indicator of the moving object detection and background subtraction methods. F-measure is defined as follows.
\begin{equation} \label{F-Measure}
\text{F-measure}  = 2* \frac {\text{Recall}*\text{Precision}}{\text{Recall}+\text{Precision}}
\end{equation}
To ensure consistency with existing methods, all the evaluation metrics are computed as defined in \cite{wang2014cdnet}.

\subsection{DBSGen Results}
\begin{figure*}[ht!]
\centering
\includegraphics[width=5.6in]{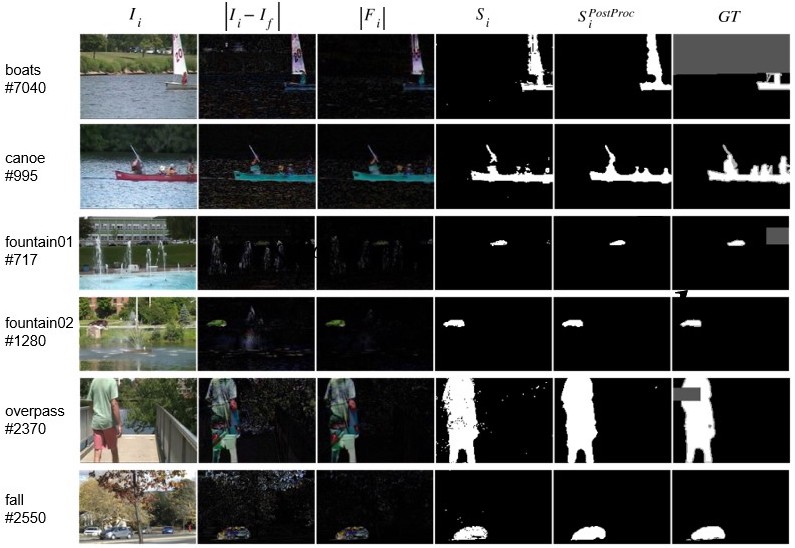}
\caption{Qualitative results of consecutive steps of DBSGen. In each row, columns from left to right show an input frame of a video, difference of the input frame and the fixed image, the obtained foreground, the binary segmented result, the post-processed segmented result, and the ground-truth.}
\label{fig:qualitative}
\end{figure*}

Qualitative results of DBSGen can be observed in Fig. \ref{fig:qualitative}. Each row shows the intermediate and final results for one frame of each video. Columns show input frames, difference of the input frames and the fixed image, the obtained foreground images, the binary segmented results, the post-processed segmented results and ground-truths, successively. Comparison between the second and third columns illustrates DBSGen was able to remove dynamic background noise to an acceptable level, before pixel-wise thresholding. Additionally, the pre-post-processing results, in the fourth column, demonstrate that DBSGen, even without the help of post-processing operations, is capable of handling dynamic background challenge to a good extent by its pixel-wise distance threshold, $R(x)$, based on dynamic entropy map, $C(x)$. The final results, in the fifth column, show DBSGen eliminates dynamic background noise successfully.

To evaluate effectiveness of the motion estimation component of DBSGen, we omitted GMCN and $L_{motion}$ that are responsible for removing some dynamic background motions by warping. The obtained results, reported in Table \ref{table:motion_comparison} in terms of FM, indicate motion estimation component plays an important role in our method and positively affects the performance of DBSGen. Table \ref{table:motion_comparison} also includes results with and without post-processing as reference points. Comparison between the second and fourth rows, where motion component is not removed, proves DBSGen's performance without post-processing step dose not drop drastically .

\begin{table}[ht!]
\caption{Performance comparison of DBSGen with or without motion estimation component, as well as with or without post-processing (PP), based on F-measure score.}
\label{table:motion_comparison}
\begin{tabular}{llrrrrrrr}
 \hline
Motion & PP  & \multicolumn{1}{l}{fnt1} & \multicolumn{1}{l}{fnt2} & \multicolumn{1}{l}{canoe} & \multicolumn{1}{l}{boats} & \multicolumn{1}{l}{over} & \multicolumn{1}{l}{fall} & \multicolumn{1}{l}{Avg.} \\
 \hline
No     & No  & 0.16                     & 0.45                     & 0.77                      & 0.34                      & 0.59                      & 0.62                     & 0.49                     \\
Yes    & No  & 0.66                     & 0.76                     & 0.86                      & 0.76                      & 0.78                      & 0.86                     & 0.78                     \\
No     & Yes & 0.27                     & 0.74                     & 0.82                      & 0.87                      & 0.79                      & 0.82                     & 0.72                     \\
Yes    & Yes & 0.73                     & 0.80                     & 0.90                      & 0.91                      & 0.87                      & 0.93                     & 0.86    \\
 \hline
\end{tabular}
\end{table}

\subsection{Comparison}
For comparison, we chose the top $30$ methods which had the best performance in terms of F-measure on Dynamic Background category of CDnet 2014 challenge results \cite{wang2014cdnet} listed on ChangeDetection.net website. The supervised methods and ensemble method IUTIS, that combines several algorithms, \cite{bianco2017combination} are not considered. In addition, CANDID algorithm \cite{mandal2018candid}, that was specifically proposed for dynamic background subtraction, is also considered.

The quantitative results are presented in Table \ref{table:methods_comparison}, where all methods are sorted based on their average FM over all videos, listed in the last column. DBSGen results are reported in the last row. As visible through last column, DBSGen achieves an average of $0.86$ in terms of FM and outperforms most of the top-ranked methods. It is only surpassed by FTSG \cite{wang2014static} and PAWCS \cite{st2015self} methods. In the ``fall'' video, we obtain the best performance along with FTSG.

Besides, DBSGen does not obtain very low performance in any of the videos unlike GMM methods \cite{zivkovic2004improved, stauffer1999adaptive}, KDE \cite{elgammal2000non} and SOBS\_CF \cite{maddalena2010fuzzy} that do not get satisfactory results on ``fountain01'' video.

\begin{table*}[ht!]
\caption{Performance comparison of the top-ranked methods, evaluated on CDnet 2014 Dynamic Background category, in terms of F-measure. The best performance achieved, in each column, is shown in bold.}
\centering
\label{table:methods_comparison}
\begin{tabular}{lrrrrrrr}
\hline
Methods                                                                     & \multicolumn{1}{l}{fountain01} & \multicolumn{1}{l}{fountain02} & \multicolumn{1}{l}{canoe} & \multicolumn{1}{l}{boats} & \multicolumn{1}{l}{overpass} & \multicolumn{1}{l}{fall} & \multicolumn{1}{l}{Average} \\
\hline
CL-VID \cite{lopez2018foreground}                                             & 0.05                           & 0.45                           & 0.93                      & 0.81                      & 0.85                         & 0.23                     & 0.55                        \\
C-EFIC \cite{allebosch2015c}                                                  & 0.27                           & 0.34                           & 0.93                      & 0.37                      & 0.90                         & 0.56                     & 0.56                        \\
EFIC \cite{allebosch2015efic}                                                 & 0.23                           & 0.91                           & 0.36                      & 0.36                      & 0.88                         & 0.72                     & 0.58                        \\
Multiscale ST BG Model \cite{lu2014multiscale}                   & 0.14                           & 0.82                           & 0.48                      & 0.89                      & 0.84                         & 0.41                     & 0.60                        \\
KDE \cite{elgammal2000non}                                         & 0.11                           & 0.82                           & 0.88                      & 0.63                      & 0.82                         & 0.31                     & 0.60                        \\
CP3-online \cite{liang2015co}                                                 & 0.54                           & 0.91                           & 0.63                      & 0.17                      & 0.64                         & 0.77                     & 0.61                        \\
DCB \cite{krungkaew2016foreground}                                            & 0.40                           & 0.83                           & 0.45                      & 0.87                      & 0.83                         & 0.30                     & 0.61                        \\
GMM\_Zivkovic \cite{zivkovic2004improved}                                     & 0.08                           & 0.79                           & 0.89                      & 0.75                      & 0.87                         & 0.42                     & 0.63                        \\
GMM\_Stauffer-Grimson \cite{stauffer1999adaptive}                         & 0.08                           & 0.80                           & 0.88                      & 0.73                      & 0.87                         & 0.44                     & 0.63                        \\
SOBS\_CF \cite{maddalena2010fuzzy}                                            & 0.11                           & 0.83                           & \textbf{0.95}             & 0.91                      & 0.85                         & 0.26                     & 0.65                        \\
SC\_SOBS \cite{maddalena2012sobs}                                             & 0.12                           & 0.89                           & \textbf{0.95}             & 0.90                      & 0.88                         & 0.28                     & 0.67                        \\
AAPSA \cite{ramirez2016auto}                                                  & 0.44                           & 0.36                           & 0.89                      & 0.76                      & 0.82                         & 0.75                     & 0.67                        \\
M4CD Version 2.0 \cite{wang2018m}                                             & 0.17                           & 0.93                           & 0.61                      & \textbf{0.95}             & 0.95                         & 0.50                     & 0.69                        \\
RMoG  \cite{varadarajan2013spatial}        & 0.20                           & 0.87                           & 0.94                      & 0.83                      & 0.90                         & 0.67                     & 0.74                        \\
WeSamBE \cite{jiang2017wesambe}                                               & 0.73                           & 0.94                           & 0.61                      & 0.64                      & 0.72                         & 0.81                     & 0.74                        \\
Spectral-360 \cite{sedky2014spectral}                                         & 0.47                           & 0.92                           & 0.88                      & 0.69                      & 0.81                         & 0.90                     & 0.78                        \\
MBS Version 0\cite{sajid2015background}& 0.52                           & 0.92                           & 0.93                      & 0.90                      & 0.90                         & 0.57                     & 0.79                        \\
MBS \cite{sajid2017universal}                    & 0.52                           & 0.92                           & 0.93                      & 0.90                      & 0.90                         & 0.57                     & 0.79                        \\
BMOG \cite{martins2017bmog}                                                   & 0.38                           & 0.93                           & \textbf{0.95}             & 0.84                      & \textbf{0.96}                & 0.69                     & 0.79                        \\
CANDID \cite{mandal2018candid}                                                & 0.55                           & 0.92                           & 0.91                      & 0.67                      & 0.92                         & 0.81                     & 0.80                        \\
SBBS \cite{varghese2017sample}           & 0.73                           & 0.93                           & 0.49                      & 0.94                      & 0.91                         & 0.88                     & 0.81                        \\
SuBSENSE \cite{st2014subsense}                                                & 0.75                           & 0.94                           & 0.79                      & 0.69                      & 0.86                         & 0.87                     & 0.82                        \\
SharedModel \cite{chen2015learning}                                           & 0.78                           & 0.94                           & 0.62                      & 0.88                      & 0.82                         & 0.89                     & 0.82                        \\
CwisarDH \cite{de2014change}                                                  & 0.61                           & 0.93                           & 0.94                      & 0.84                      & 0.90                         & 0.75                     & 0.83                        \\
WisenetMD \cite{lee2019wisenetmd}                                             & 0.75                           & \textbf{0.95}                  & 0.87                      & 0.71                      & 0.87                         & 0.87                     & 0.84                        \\
AMBER \cite{wang2014fast}                                                     & 0.77                           & 0.93                           & 0.93                      & 0.85                      & 0.95                         & 0.63                     & 0.84                        \\
CwisarDRP \cite{de2017wisardrp}                                               & 0.69                           & 0.92                           & 0.91                      & 0.84                      & 0.92                         & 0.82                     & 0.85                        \\
CVABS \cite{icsik2019cvabs}                                                   & 0.77                           & 0.94                           & 0.88                      & 0.81                      & 0.86                         & 0.91                     & 0.86                        \\
SWCD \cite{isik2018swcd}                                                      & 0.76                           & 0.93                           & 0.92                      & 0.85                      & 0.85                         & 0.88                     & 0.86                        \\
FTSG \cite{wang2014static}           & \textbf{0.81}                  & \textbf{0.95}                  & 0.69                      & \textbf{0.95}             & 0.94                         & \textbf{0.93}            & 0.88                        \\
PAWCS \cite{st2015self}                                                       & 0.78                           & 0.94                           & 0.94                      & 0.84                      & \textbf{0.96}                & 0.91                     & \textbf{0.89}               \\
\textbf{DBSGen}                                                             & 0.73                           & 0.80                           & 0.90                      & 0.91                      & 0.87                         & \textbf{0.93}            & 0.86     \\
\hline
\end{tabular}
\end{table*}

\section{Conclusion} \label{sec:Conclusion}
We have presented a generative neural net based background subtraction method called DBSGen to handle dynamic background challenge. DBSGen is unsupervised, so it does not need annotated ground-truth data for training, furthermore, it gets optimized in an end-to-end way. Besides, it has a minimal post-processing step, which can be also omitted without a significant performance drop. 
DBSGen estimates a dense dynamic motion map by use of a Generative Multi-resolution Convolutional Network (GMCN) and warps the input images by the obtained motion map. Then, a Generative Fully Connected Network (GFCN) generates background images by using warped input images in its reconstruction loss term. In the following step, a pixel-wise distance threshold that utilizes a dynamic entropy map obtains the binary segmented results. Finally, a basic median filter and morphological closing is applied as the post-processing step.
Experiments on Dynamic Background category of CDnet 2014 demonstrates that DBSGen surpasses all previously tested methods, which are unsupervised and not ensemble of several methods, on CDnet 2014 in terms of F-measure. Only two state-of-the-art methods outperform DBSGen. Overall, quantitative and qualitative results confirm that DBSGen is capable of eliminating dynamic background motions quite effectively.

For the future work, we want to merge our previous framework, NUMOD \cite{bahri2018online}, that can cope with illumination changes and shadows, with DBSGen. Also, we want to consider some advanced post-processing techniques to improve the results.

\bibliographystyle{IEEEtran}
\bibliography{ref}

\end{document}